\documentclass[11pt]{article}
\begin{document}

\title{\bf\large Response to `Comment on ``Experimental demonstration of a universally valid error-disturbance uncertainty relation in spin measurements''\ ' }
\author{\normalsize Jacqueline Erhart,
Stephan Sponar,
Georg Sulyok,
Gerald Badurek, \\
Masanao Ozawa, and
Yuji Hasegawa}
\date{}
\newcommand{\bracket}[1]{\langle #1 \rangle}
\maketitle

The author of the comment \cite{comment} claimed that the experimental results for a universally valid uncertainty relation
cannot be recognized to establish a violation of the Heisenberg-type uncertainty relation.
However, after careful examination of the author's argument, we regard the author's argument to be on an improper basis, thus disagree with his opinion. Hereby, we provide arguments to disprove his objection.

\sloppy
According to his statement ``To insist the violation of the Heisenberg-type uncertainty relation,
$\epsilon(Q)=0$ or $\eta(P)=0$, one has to prove $|\bracket{[N(Q),D(P)]}|=0$,''
he appears to formulate the violation to be the case where $\epsilon(Q)=0$ or $\eta(P)=0$.
However, in our paper the violation is taken to be the cases where the relation
$\epsilon(A)\eta(B)\ge\frac{1}{2}|\bracket{[A,B]}|$
does not hold, and we observed the violation for all the parameter
values of $\phi$ actually tested.

The author conclude the comments by writing ``In conclusion, the experimental set-up of ref. \cite{Erhard12} is not satisfied
the necessary condition to be true for the UVUR proved by Ozawa in \cite{Ozawa03PLA, Ozawa03PRA}, because the neutron
spin measurement (``projective measurement'' as well) does not have the unitary operator to defined
the equal-time commutation relation''.

In contrast to what the author claimed, the projective measurement is defined by a family of projection
operators $M_m$, instead of a unitary operator $U$, and the root-mean-square error is defined by equation (5)
in Ref. \cite{Erhard12} using the projection operators $M_m$, instead of using the unitary operator $U$.
It is well-known that if the measurement is defined by a unitary operator $U$, then the
definition of the root-mean-square error using the unitary operator $U$ and the definition
(5) in Ref. \cite{Erhard12} using only projection operators $M_m$ are equivalent.
A detailed account on those equivalences between definitions are given in
[M. Ozawa, Uncertainty relations for noise and disturbance in generalized
quantum measurements, Ann. Phys. (N.Y.) 311, 350-416  (2004)].
Thus, the author's criticism based on the absence of the unitary operator is also irrelevant.

\end{document}